\documentclass[conference]{IEEEtran}
\usepackage{graphicx}
\usepackage{array}

\usepackage{multirow}
\usepackage{longtable}
\usepackage{float}
\usepackage[justification=centering]{caption}
\usepackage{booktabs}
\begin{document}
\title{On Which Skills do Indian Universities Evaluate Software Engineering
Students?}
\author{\IEEEauthorblockN{Hansaraj S. Wankhede, Sanil S. Gandhi, Arvind W Kiwelekar}
\IEEEauthorblockA{Department of Computer Engineering\\
Dr. Babasaheb Ambedkar Technological University\\
Lonere, Maharashtra, India \\
Email: \{hswankhede@dbatu.ac.in, sanil.gandhi15@gmail.com, awk@dbatu.ac.in\}}
}
\maketitle
\begin{abstract}
Universities conduct examinations to evaluate  acquired skills  and knowledge
gained by  students.
An assessment of skills and knowledge levels evaluated during Software Engineering examinations 
is presented in this paper.
The question items asked during examinations  are analyzed from three dimensions that are
cognitive levels, knowledge levels and knowledge areas. 
The  Revised Bloom's Taxonomy is used to classify question items 
along the dimensions of cognitive levels and knowledge levels. 
Question items are also
classified in various  knowledge areas specified
in ACM/IEEE's Computer Science Curricula. The analysis presented
in this paper will be useful for software engineering educators to
devise corrective interventions and employers of fresh graduates 
to design pre-induction training programs.
\end{abstract}
\IEEEpeerreviewmaketitle

\section{Introduction}
An assessment of the skills acquired and knowledge gained through a course on
Software Engineering is equally useful to academicians as well as industry
professionals. Academicians can use the  results of the assessment to devise
appropriate interventions in case of the assessment results do not conform to
the set learning objectives. Employers of fresh graduates may use the results of
the assessment to design pre-induction training programs.

One way to perform such an assessment is to analyze 
question papers used for  conducting  end-semester examinations because
it includes  the  most relevant  information 
required for such an assessment. An end-semester question paper is typically 
designed to test students on diverse range of skills such as \textit{to recall}
a learned topic or  \textit{to apply} a learned method to solve a  particular
problem. Further  question papers include  questions from all the knowledge
areas that are expected to be covered in a course on Software Engineering. 

In this paper we classify  questions asked in an examination  along
three dimensions, namely, cognitive levels, knowledge levels and knowledge areas. 
The categories included in the Revised Bloom's Taxonomy \cite{krathwohl2002revision} are used to
classify  question items along the dimensions of  knowledge and cognitive
levels.  Question items are also classified according to the topics included under various knowledge areas of Software Engineering
defined in  ACM/IEEE's Computer Science Curricula 2013 \cite{ACMCurricula,CS2013}.

\section{Analysis framework}
\begin{figure}[]
\begin{center}
\includegraphics[scale=0.4]{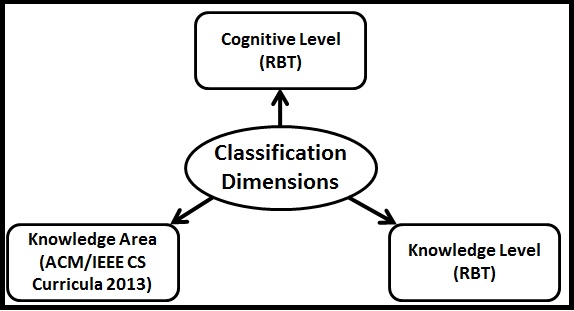} 
\caption{Three domains in the Analysis Framework}\label{f1}
\label{ClassDim}
\end{center}
\end{figure}

The classification framework used to analyze the question items is derived from
two different sources. The main intention of the classification  framework is to
analyze question items from three different dimensions as shown in Figure \ref{f1}. 
The first two dimensions
are cognitive levels, knowledge levels as defined in  Revised Bloom's Taxonomy
(RBT) \cite{krathwohl2002revision}. 
Further, each question item asked in
Software Engineering examinations belongs to a particular topic or a course
unit. Hence the topics that are covered under the Software Engineering knowledge
areas of  ACM/IEEE Computer Science Curricula 2013 are also included. The first
two dimensions cover generic learning skills that educators intend to impart in
students while the third dimension covers domain specific skills that employers
expect from a fresh-graduate.

\begin{figure}[]
\begin{center}
\includegraphics[scale=0.5]{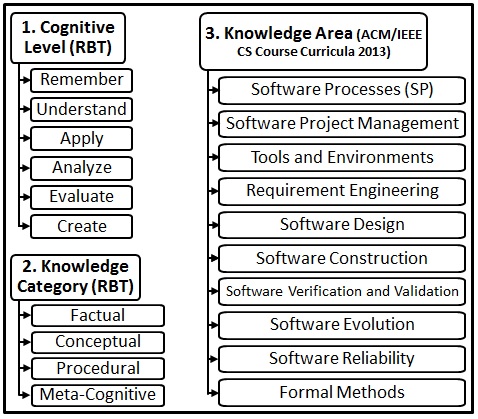}
\caption{Classification Categories in the Framework}\label{f2}
\end{center}
\label{DimDetail}
\end{figure}
\subsection{Cognitive Levels}

The cognitive process dimension in RBT is broadly organized in six different
categories namely  Remember, Understand,   Apply, Analyze, Evaluate and Create as shown in Figure \ref{f2}. 
The category {\em Remember} captures the activity of retrieving knowledge from
long-term memory.  The activities of recognizing and recalling information,
objects and events belong to the {\em Remember} category. The category {\em
Understand} means to construct the meaning out of the  learning material
presented in the form of either lectures or notes.  The acts of interpreting,
exemplifying, classifying, summarizing, inferring, comparing, and explaining
belong to the category {\em Understand}. The third category {\em Apply} refers
to carry out or use a procedure in a given situation and it includes the acts of
executing and implementing. 
The fourth category of {\em Analyze} refers to breaking down the learning
material into its part and to relating parts to establish overall structure. The
acts of differentiating, organizing and attributing are considered as analytic
processes.   The fifth category {\em Evaluate} means the acts of checking and
making judgments based on some criteria. The last cognitive process category
from RBT is {\em Create} and it means the acts of generating, planning and
producing 
some product.
 Few example questions and their mapped cognitive levels
are shown in Table \ref{cclass}.

\begin{table}[]
\centering
\caption{Question mapping to Cognitive levels in RBT} \label{cclass}
\begin{tabular}{|m{1cm}| m{6cm} |} \hline
   
  \begin{center}
  \textbf{Cognitive Level}
  \end{center} 
  &
  \begin{center}
    \textbf{Question Item}
    \end{center}
    \\ \hline
    Remember & What is quality assurance ? What are different parameters of
quality?
     \\ \hline
    Understand & Explain incremental process model with a suitable example.
     \\ \hline
    Apply & How do you calculate Function Point(FP) and how it is used in
estimation of a software project?
     \\ \hline
    Analyze & Discuss and Compare Test driven development (TDD) and Design driven
testing (DDT).  \\ \hline
    Evaluate & Which life cycle model would you follow for developing following
project and why.  (a) Library management system (b) A web application 
    \\ \hline
    Create & Develop a complete Use Case System for ATM machine.
    \\ \hline
      \end{tabular}
      \end{table}

\subsection{Knowledge Levels} \label{ka}

As shown in Figure \ref{f2}, the knowledge dimension in RBT is classified into four categories of {\em
Factual, Conceptual, Procedural} and {\em Meta-cognitive}
knowledge.  The factual information about specific terminologies (e.g.,
Products, Processes, Life Cycle models ) and basic elements that students must be well versed
with are captured under the factual knowledge. The conceptual knowledge 
category includes the knowledge about classification categories, principles,
models and theories.  Some examples of conceptual knowledge 
are knowledge about life cycle models, and principle of  modularity. The
knowledge about procedures, methods, algorithms are included under the category
of  {\em Procedural Knowledge}. An example of procedural knowledge is 
methods for Object-Oriented Analysis (e.g., CRC Card).   The last category {\em meta-cognitive}
knowledge corresponds to knowledge
about cognition itself and understanding one's own cognitive abilities.
Table \ref{cq} depicts the example questions and mapped knowledge level categories.
\begin{table}[]
\centering
\caption{Question mapping to Knowledge Category levels in RBT} \label{cq}
\begin{tabular}{|m{2.5cm}| m{5cm} |} \hline
   
  \begin{center}
  \textbf{Knowledge Category}
  \end{center} 
  &
  \begin{center}
    \textbf{Question Item}
    \end{center}
    \\ \hline
    Factual Knowledge & Explain the difference between software and hardware
characteristics.
    \\ \hline
    Conceptual Knowledge & Explain the waterfall life cycle model.
     \\ \hline
  Procedural Knowledge &  Explain how project scheduling and tracking is done
for a software development project ? 
  \\ \hline
    Meta-cognitive Knowledge & No question is mapped to this category  \\ \hline
     \end{tabular}
      \end{table}

\subsection{Software Engineering Knowledge Area}

A set of guidelines are specified in ACM/IEEE's Computer Science curricula
(CS2013)\cite{CS2013} to design a undergraduate program in Computer Science. In
CS2013, the knowledge body of Computer Science is organized into eighteen Knowledge
Areas (KA). Each  KA is further sub divided
in various Knowledge Units (KU). Software Engineering is one of the eighteen KAs
which is further subdivided into ten different KUs as shown in Figure \ref{f2}.
In this paper, we have selected the CS2013 as a reference
knowledge body with an intention to bridge the non-uniformity in course content
of the courses on Software Engineering offered by various Indian Universities.

\begin{table}[]
\centering
\caption{Question item mapping to Knowledge Units of Knowledge Area in
ACM/IEEE's Computer Science Curricula 2013}
\begin{tabular}{|m{3cm}| m{5cm} |} \hline
   
  \begin{center}
  \textbf{Knowledge Unit}
  \end{center} 
  &
  \begin{center}
    \textbf{Question Item}
    \end{center}
    \\ \hline
    Software Processes (SP) & Compare waterfall model and spiral model.
     \\ \hline
    Software Project Management (SPM) & Describe project scheduling and tracking
with any suitable example.
    \\ \hline
    Tools and Environments (TE) & Explain Software Configuration Management in
detail. 
    \\ \hline
    Requirement Engineering (RE) & Explain different steps in requirement
engineering.
    \\ \hline
    Software Design (SD) &  List and explain the fundamental concepts for
software design.
     \\ \hline
    Software Construction (SC) & Compare conventional approach and object
oriented approach to software development ? What are the advantages of OOAD ?
    \\ \hline
    Software Verification and \newline Validation(SVV) & What is software
testing ? Explain the software testing strategies.
    \\ \hline
    Software Evolution (SE) & Define "Program Evolution Dynamics". Discuss the
Lehman laws for program evolution dynamics.
    \\ \hline
    Software Reliability (SR) &  What do you understand by software reliability
? 
     \\ \hline
    Formal Methods (FM) & No question is mapped to this topic \\ \hline
      \end{tabular}
      \end{table}

\begin{table}[t]
 \centering
 \caption{Question Paper Data Collection} \label{universities}
 \begin{tabular}{|m{0.3cm}|m{4.5cm}|m{0.75cm}|m{0.75cm}|} \hline  
  \begin{center}\textbf{Sr. No.}\end{center}
    &    
  \begin{center}\textbf{University}\end{center}
    &
  \begin{center}\textbf{No. of Question Papers}\end{center}
    &
  \begin{center}\textbf{No. of Question Items}\end{center}   \\ \hline
   1. & Viswesarayya Technological University (VTU) & 7 & 146 \\ \hline 
   2. & Savitribai Phule Pune University (PU) & 6 & 174 \\ \hline
   3. & Mumbai University (MU) & 7 & 94 \\ \hline
   4. & Gujarat Technological University (GTU)& 6 & 103 \\ \hline
   5. & Anna University (AU) & 4 & 103 \\ \hline
   6. & West Bengal Technological University (WBTU) & 3 & 68 \\ \hline
   7. & Punjab Technological University  (PTU)& 3 & 57 \\ \hline
   8. & Dr. Babasaheb Ambedkar Technological \newline University  (DBATU) & 3 & 49 \\
\hline  
\end{tabular}
\end{table}

\begin{table}[]
\centering
\caption{Action Verbs \cite{RBTverb} for Cognitive Categories in RBT} \label{actionverbs}
\begin{tabular}{|m{1cm}| m{7cm} |} \hline
   
  \begin{center}
  \textbf{Category}
  \end{center} 
     &
    \begin{center}
      \textbf{Action Verbs}
      \end{center}  \\ \hline
 Remember & Choose, Define, Find, How, Label, List, Match, Name, Omit, Recall,
Relate, Select, Show, Spell, Tell, What, When, Where, Which, Who, Why \\ \hline
 Understand & Classify, Compare, Contrast, Demonstrate, Explain, Extend,
Illustrate, Infer, Interpret, Outline, Relate, Rephrase, Show, Summarize,
Translate \\ \hline
 Apply & Apply, Build, Choose, Construct, Develop, Experiment with, Identify,
Interview, Make use of, Model, Organize, Plan, Select, Solve, Utilize \\ \hline
 Analyze & Analyze, Assume, Categorize, Classify, Compare, Conclusion, Contrast,
Discover, Dissect, Distinguish, Divide, Examine, Function, Inference, Inspect,
List, Motive, Relationships, Simplify, Survey, Take part in, Test for, Theme \\
\hline
 Evaluate & Agree, Appraise, Assess, Award, Choose, Compare, Conclude, Criteria,
Criticize, Decide, Deduct, Defend, Determine, Disprove, Estimate, Evaluate,
Explain, Importance, Influence, Interpret, Judge, Justify, Mark, Measure,
Opinion, Perceive, Prioritize, Prove, Rate, Recommend, Rule on, Select, Support,
Value\\ \hline
 Create & Adapt, Build, Change, Choose, Combine, Compile, Compose,Construct,
Create, Delete, Design, Develop, Discuss, Elaborate, Estimate, Formulate,
Happen, Imagine, Improve, Invent, Make up, Maximize, Minimize, Modify, Original,
Originate, Plan, Predict, Propose, Solution, Solve, Suppose, Test, Theory \\
\hline
\end{tabular}
\end{table}

\section{The Analysis Method}

The analysis of  question papers  is carried with an intention to
answer the following  questions.

(i) \textit{Do SE examinations test student  for all cognitive skills? }\newline 
This question is significant because
students of an engineering under-graduate  program are expected to be 
evaluated on higher order thinking skills such as {\em Analysis} and {\em Synthesis}
rather than evaluating them on skills such as {\em Remember} and {\em Understand}.

(ii) \textit{For which kinds of knowledge students are tested during SE examinations}?\newline
Answering this question is important because certain courses contain a specific
kind of knowledge.  For example, the content of a course on  Data  Structure and
Algorithms is of  \textit{Procedural} type while the majority of  the contents
of a course on Software Engineering is of {\em Conceptual} type. The  question items
asked in an examination should reflect this tacit assumption.

 (iii) \textit{Do SE examinations give sufficient coverage to all the knowledge
units?}\newline This question is significant to answer because it verifies that
an examination sufficiently covers all the knowledge units that are expected to
be covered or whether it is skewed towards a particular topic.  

\subsection{Activities}
Majority of Indian Universities adopt the \textit{standardized test} as an
assessment method to test the knowledge and skills  of  the student enrolled for
a particular course at an affiliated college. In standardized test, examinations
are administered by  universities. Students  appear for an examination and
they answer the  same question paper delivered at the same time across various
affiliated colleges. The survey presented in this paper considers only those
institutes which adopt  \textit{standardized test} as an assessment method. 
Autonomous colleges conducting examinations specific to students enrolled
in one particular institute are excluded from the survey.  This section describes
the main activities conducted during the survey.

\subsubsection{Collection of Question papers}

Question items are the basic unit of analysis for the survey
presented in this paper. Questions items are collected from
end-semester examinations conducted by various universities. Most of the 
universities offer a course on Software Engineering 
during  third year under-graduate programme in either Computer Engineering or
Information Technology.  Few   universities  from all the four regions
of India are selected for the analysis.  Easy accessibility of the question
papers in public domain   is the main criteria for selecting the universities.
Most of the question papers included in the survey are downloaded from
the official web sites of the  universities. Some of the question papers 
are also downloaded from the web-sites\cite{paper1,paper2} hosting  study material
for engineering students.
Question papers for the examinations held during last five years are used for the analysis.
Table \ref{universities} shows the details of the
number of question papers selected and the total number of question items from
the respective university.

\subsubsection{Preparation of Question Bank}

A question bank  in the form of a spreadsheet is  prepared by picking up question items from the  selected question papers.
For each question item, the information
about text of a  question, name of a  university,  examination year, assigned categories i.e. knowledge,
 cognitive, and knowledge area is stored. 
The question bank  includes about eight hundred questions asked in about
forty question papers.  Some of the question items are duplicated and not stored in the question bank 
because the same questions
may be repeated in  multiple examinations.

\subsubsection{Assignment of Categories}

Each question item is classified into three different categories
i.e. cognitive, knowledge type and knowledge units.
To assign a cognitive category,  the  category wise list of
action verbs prepared by Azusa Pacific University, California \cite{RBTverb}
shown in Table \ref{actionverbs} is used.   
Table \ref{cclass} shows the  assignment of cognitive categories to few question items.
The knowledge category is assigned by interpreting noun-phrase in the question
item.  The guidelines specified in \cite{krathwohl2002revision} are used  to classify the question items
in various knowledge categories. 
Some of the guidelines used during interpretation are also described in Section \ref{ka}.
The knowledge unit is assigned to a question item by interpreting the content of
the question. 
About eight hundred question items are analyzed and  categories are assigned
from three different perspectives. A tool has been implemented in Java to  assign
cognitive level categories.
Initially cognitive level categorization is   manually performed by the first author which
has taught a course on Software Engineering.  
Question items are also categorized through a tool and 
verified by the second and third authors.

\begin{table}[]
 \centering
 \caption{Paper wise Analysis Cognitive Categorization  (\% distribution) \label{ca}}
\renewcommand{\arraystretch}{1.7}
 \begin{tabular}{|m{1.5cm}|m{0.6cm}|m{0.6cm}|m{0.6cm}|m{0.6cm}|m{0.6cm}|m{0.6cm}
|} \hline
  PaperID & Reme-mber & Under-stand & Apply & Analy-ze & Evalu-ate & Create  \\
[2.5pt]   \hline
 MU2014S & 6.25	& 43.75	& 18.75 & 0.00 & 25.00 & 6.25  \\ \hline
 MU2013S & 0.00	& 61.54	& 7.69 & 0.00 & 15.38 & 15.38 \\ \hline
 MU2012S & 0.00 & 72.73 & 9.09 & 0.00 & 0.00 & 18.18 \\ \hline
 MU2014W & 26.67 & 53.33 & 13.33 & 6.67 & 0.00 & 0.00\\ \hline
 MU2013W & 0.00 & 38.46 & 23.08 & 23.08 & 15.38 & 0.00\\ \hline
 MU2012W & 7.69 & 76.92 & 7.69 & 0.00 & 0.00 & 7.69\\ \hline
 MU2011W & 0.00 & 69.23 & 15.38 & 7.69 & 0.00 & 7.69\\ \hline
 PU2014W & 20.00 & 66.67 & 6.67 & 3.33 & 3.33 & 0.00\\ \hline
 PU2013W & 13.33 & 76.67 & 3.33 & 3.33 & 3.33 & 0.00\\ \hline
 PU2012W & 10.00 & 60.00 & 10.00 & 10.00 & 3.33 & 6.67\\ \hline
 PU2014S & 28.57 & 60.71 & 3.57 & 3.57 & 3.57 & 0.00\\ \hline
 PU2013S & 3.57 & 53.57 & 14.29 & 10.71 & 3.57 & 14.29\\ \hline
 PU2011S & 3.57 & 64.29 & 21.43 & 3.57 & 7.14 & 0.00\\ \hline
 VU2014W & 0.00 & 76.19 & 23.81 & 0.00 & 0.00 & 0.00\\ \hline
 VU2013W & 29.41 & 58.82 & 5.88 & 0.00 & 5.88 & 0.00\\ \hline
 VU2012W & 0.00 & 85.71 & 0.00 & 14.29 & 0.00 & 0.00\\ \hline
 VU2011W & 8.00 & 80.00 & 0.00 & 0.00 & 4.00 & 8.00\\ \hline
 VU2012M & 13.04 & 78.26 & 0.00 & 8.70 & 0.00 & 0.00\\ \hline
 VU2013M & 13.64 & 77.27 & 4.55 & 0.00 & 4.55 & 0.00\\ \hline
 VU2014M & 20.83 & 58.33 & 0.00 & 0.00 & 4.17 & 16.67\\ \hline
 GU2014W & 0.00 & 70.59 & 11.76 & 5.88 & 0.00 & 11.76\\ \hline
 GU2013W & 5.56 & 61.11 & 11.11 & 11.11 & 0.00 & 11.11\\ \hline
 GU2014S & 17.65 & 58.82 & 0.00 & 17.65 & 0.00 & 5.88\\ \hline
 GU2013S & 17.65 & 52.94 & 17.65 & 5.88 & 5.88 & 0.00\\ \hline
 GU2012S & 0.00 & 41.18 & 11.76 & 11.76 & 0.00 & 35.29\\ \hline
 GU2011S & 11.76 & 70.59 & 5.88 & 0.00 & 0.00 & 11.76\\ \hline
 AU2014S & 36.00 & 52.00 & 4.00 & 0.00 & 0.00 & 8.00\\ \hline
 AU2013S & 31.03 & 62.07 & 6.90 & 0.00 & 0.00 & 0.00\\ \hline
 AU2013W & 41.38 & 27.59 & 0.00 & 10.34 & 3.45 & 17.24\\ \hline
 AU2012W & 15.00 & 55.00 & 15.00 & 0.00 & 10.00 & 5.00\\ \hline
 WBTU2013 & 33.33 & 33.33 & 11.11 & 5.56 & 11.11 & 5.56\\ \hline
 WBTU2012 & 19.35 & 51.61 & 9.68 & 3.23 & 6.45 & 9.68\\ \hline
 WBTU2011 & 36.84 & 36.84 & 5.26 & 10.53 & 0.00 & 10.53\\ \hline
 PTU2010S & 38.89  & 44.44 & 0.00 & 5.56 & 5.56 & 5.56\\ \hline
 PTU2009W & 76.19 & 23.81 & 0.00 & 0.00 & 0.00 & 0.00\\ \hline
 PTU2009S & 38.89 & 61.11 & 0.00 & 0.00 & 0.00 & 0.00\\ \hline
 DBATU2015S & 25.00 & 50.00 & 8.33 & 0.00 & 8.33 & 8.33\\ \hline
 DBATU2014W & 22.22 & 55.56 & 0.00 & 16.67 & 5.56 & 0.00\\ \hline
 DBATU2014S & 31.58 & 57.89 & 0.00 & 5.26 & 5.26 & 0.00\\ \hline 
\end{tabular}
\end{table}

\begin{table}[]
 \centering
 \caption{Paper wise Analysis Knowledge Categorization  (\% distribution) \label{kl}}
\renewcommand{\arraystretch}{1.7}
 \begin{tabular}{|m{1.5cm}|m{1cm}|m{1cm}|m{1cm}|m{1cm}|} \hline
  PaperID & Factual & Concep-tual & Proced-ural & Meta-Cognitive    \\ [2.5pt]
\hline
 MU2014S & 18.75 & 50.00 & 31.25 & 0.00 \\ \hline 
 MU2013S & 0.00 & 61.54 & 38.46 & 0.00\\ \hline 
 MU2012S & 0.00 & 63.64 & 36.36 & 0.00\\ \hline 
 MU2014W & 26.67 & 53.33 & 20.00 & 0.00\\ \hline 
 MU2013W & 7.69 & 53.85 & 38.46 & 0.00\\ \hline 
 MU2012W & 15.38 & 69.23 & 15.38 & 0.00\\ \hline 
 MU2011W & 0.00 & 53.85 & 46.15 & 0.00\\ \hline 
 PU2014W & 6.67 & 66.67 & 26.67 & 0.00\\ \hline 
 PU2013W & 6.67 & 60.00 & 33.33 & 0.00\\ \hline 
 PU2012W & 3.33 & 43.33 & 53.33 & 0.00\\ \hline 
 PU2014S & 39.29 & 25.00 & 35.71 & 0.00\\ \hline 
 PU2013S & 10.71 & 39.29 & 50.00 & 0.00\\ \hline 
 PU2011S & 7.14 & 53.57 & 39.29 & 0.00\\ \hline 
 VU2014W & 4.76 & 28.57 & 66.67 & 0.00\\ \hline 
 VU2013W & 17.65 & 47.06 & 35.29 & 0.00\\ \hline 
 VU2012W & 14.29 & 35.71 & 50.00 & 0.00\\ \hline 
 VU2011W & 24.00 & 20.00 & 56.00 & 0.00\\ \hline 
 VU2012M & 26.09 & 43.48 & 30.43 & 0.00\\ \hline 
 VU2013M & 40.91 & 22.73 & 36.36 & 0.00\\ \hline 
 VU2014M & 25.00 & 33.33 & 41.67 & 0.00\\ \hline 
 GU2014W & 5.88 & 47.06 & 47.06 & 0.00\\ \hline 
 GU2013W & 11.11 & 50.00 & 38.89 & 0.00\\ \hline 
 GU2014S & 41.18 & 17.65 & 41.18 & 0.00\\ \hline 
 GU2013S & 17.65 & 58.82 & 23.53 & 0.00\\ \hline 
 GU2012S & 11.76 & 35.29 & 52.94 & 0.00\\ \hline 
 GU2011S & 35.29 & 41.18 & 23.53 & 0.00\\ \hline 
 AU2014S & 40.00 & 36.00 & 24.00 & 0.00\\ \hline 
 AU2013S & 34.48 & 27.59 & 37.93 & 0.00\\ \hline 
 AU2013W & 34.48 & 20.69 & 44.83 & 0.00\\ \hline 
 AU2012W & 20.00 & 55.00 & 25.00 & 0.00\\ \hline 
 WBTU2013 & 38.89 & 27.78 & 33.33 & 0.00\\ \hline 
 WBTU2012 & 29.03 & 41.94 & 29.03 & 0.00\\ \hline 
 WBTU2011 & 21.05 & 31.58 & 47.37 & 0.00\\ \hline 
 PTU2010S & 50.00 & 16.67 & 33.33 & 0.00\\ \hline 
 PTU2009W & 61.90 & 14.29 & 23.81 & 0.00\\ \hline 
 PTU2009S & 16.67 & 27.78 & 55.56 & 0.00\\ \hline 
 DBATU2015S & 33.33 & 33.33 & 33.33 & 0.00\\ \hline 
 DBATU2014W & 33.33 & 44.44 & 22.22 & 0.00\\ \hline 
 DBATU2014S & 21.05 & 63.16 & 15.79 & 0.00\\ \hline   
  
\end{tabular}
\end{table}
\begin{table}[]
 \centering
 \caption{Paper wise Analysis for Knowledge Areas  (\% distribution) \label{ku}}
\renewcommand{\arraystretch}{1.7} 
\setlength\tabcolsep{2pt}
 \begin{tabular}{|m{1.2cm}|m{0.6cm}|m{0.6cm}|m{0.6cm}|m{0.6cm}|m{0.6cm}|m{
0.6cm}|m{0.6cm}|m{0.6cm}|m{0.6cm}|m{0.5cm}|} \hline
  PaperID& SP& SPM& TE& RE& SD& SC& SVV& SE& SR& FM \\ \hline
  MU2014S& 25.00 & 31.25& 12.50& 6.25& 25.00& 0.00& 0.00& 0.00& 0.00& 0.00 \\
\hline
  MU2013S & 0.00 & 25.00 & 25.00 & 16.67 & 16.67 & 0.00 & 16.67 & 0.00 & 0.00 &
0.00 \\ \hline
  MU2012S & 0.00 & 33.33 & 11.11 & 	22.22 & 33.33 & 0.00 & 0.00 & 0.00 &
0.00 & 0.00 \\ \hline
  MU2014W & 14.29 & 28.57 & 7.14 & 14.29 & 14.29 & 7.14 & 7.14 & 7.14 & 0.00 &
0.00 \\ \hline
  MU2013W & 8.33 & 33.33 & 16.67 & 8.33 & 8.33 & 16.67 & 8.33 & 0.00 & 0.00 &
0.00 \\ \hline
  MU2012W & 8.33 & 25.00 & 8.33 & 25.00 & 25.00 & 0.00 & 8.33 & 0.00 & 0.00 &
0.00 \\ \hline
  MU2011W & 0.00 & 33.33 & 8.33 & 16.67 & 16.67 & 8.33 & 8.33 & 0.00 & 8.33 &
0.00 \\ \hline
  PU2014W & 20.00 & 30.00 & 0.00 & 0.00 & 30.00 & 0.00 & 20.00 & 0.00 & 0.00 &
0.00 \\ \hline
  PU2013W & 23.33 & 23.33 & 3.33 & 3.33 & 26.67 & 0.00 & 20.00 & 0.00 & 0.00 &
0.00 \\ \hline
  PU2012W & 23.33 & 23.33 & 3.33 & 3.33 & 26.67 & 0.00 & 20.00 & 0.00 & 0.00 &
0.00 \\ \hline
  PU2014S & 25.00 & 21.43 & 7.14 & 10.71 & 14.29 & 0.00 & 14.29 & 3.57 & 3.57 &
0.00 \\ \hline
  PU2013S & 14.29 & 28.57 & 3.57 & 10.71 & 17.86 & 3.57 & 21.43 & 0.00 & 0.00 &
0.00 \\ \hline
  PU2011S & 17.86 & 32.14 & 3.57 & 3.57 & 28.57 & 0.00 & 14.29 & 0.00 & 0.00 &
0.00 \\ \hline
  VU2014W & 25.00 & 5.00 & 0.00 & 20.00 & 25.00 & 0.00 & 15.00 & 5.00 & 5.00 &
0.00 \\ \hline
  VU2013W & 23.53 & 17.65 & 0.00 & 17.65 & 17.65 & 5.88 & 5.88 & 5.88 & 5.88 &
0.00 \\ \hline
  VU2012W & 28.57 & 14.29 & 0.00 & 14.29 & 7.14 & 0.00 & 14.29 & 7.14 & 14.29 &
0.00 \\ \hline
  VU2011W & 20.00 & 16.00 & 8.00 & 16.00 & 24.00 & 4.00 & 0.00 & 4.00 & 8.00 &
0.00 \\ \hline
  VU2012M & 26.09 & 17.39 & 0.00 & 17.39 & 13.04 & 0.00 & 13.04 & 	0.00 &
13.04 & 0.00 \\ \hline
  VU2013M & 22.73 & 22.73 & 0.00 & 13.64 & 18.18 & 0.00 & 9.09 & 4.55 & 9.09 &
0.00 \\ \hline
  VU2014M & 25.00 & 12.50 & 0.00 & 12.50 & 25.00 & 4.17 & 8.33 & 0.00 & 12.50 &
0.00 \\ \hline
  GU2014W & 35.29 & 17.65 & 5.88 & 11.76 & 17.65 & 0.00 & 11.76 & 0.00 & 0.00 &
0.00 \\ \hline
  GU2013W & 22.22 & 27.78 & 11.11 & 0.00 & 22.22 & 0.00 & 16.67 & 0.00 & 0.00 &
0.00 \\ \hline
  GU2014S & 17.65 & 23.53 & 5.88 & 29.41 & 11.76 & 0.00 & 11.76 & 0.00 & 0.00 &
0.00 \\ \hline
  GU2013S & 18.75 & 31.25 & 6.25 & 6.25 & 18.75 & 0.00 & 12.50 & 0.00 & 6.25 &
0.00 \\ \hline
  GU2012S & 29.41 & 11.76 & 0.00 & 5.88 & 35.29 & 0.00 & 17.65 & 0.00 & 0.00 &
0.00 \\ \hline
  GU2011S & 29.41 & 17.65 & 5.88 & 5.88 & 11.76 & 5.88 & 17.65 & 5.88 & 0.00 &
0.00 \\ \hline
  AU2014S & 24.00 & 12.00 & 8.00 & 12.00 & 20.00 & 4.00 & 20.00 & 0.00 & 0.00 &
0.00 \\ \hline
  AU2013S & 6.90 & 17.24 & 6.90 & 13.79 & 24.14 & 6.90 & 20.69 & 0.00 & 3.45 &
0.00 \\ \hline
  AU2013W & 13.79 & 17.24 & 3.45 & 24.14 & 17.24 & 3.45 & 20.69 & 0.00 & 0.00 &
0.00 \\ \hline
  AU2012W & 21.05 & 15.79 & 0.00 & 21.05 & 21.05 & 0.00 & 21.05 & 0.00 & 0.00 &
0.00 \\ \hline
  WBTU2013 & 18.75 & 18.75 & 6.25 & 12.50 & 25.00 & 6.25 & 6.25 & 0.00 & 6.25 &
0.00 \\ \hline
  WBTU2012 & 3.33 & 16.67 & 0.00 & 13.33 & 23.33 & 3.33 & 30.00 & 3.33 & 6.67 &
0.00 \\ \hline
  WBTU2011 & 5.88 & 23.53 & 17.65 & 5.88 & 5.88 & 0.00 & 29.41 & 0.00 & 11.76 &
0.00 \\ \hline
  PTU2010S & 11.11 & 22.22 & 0.00 & 22.22 & 5.56 & 0.00 & 33.33 & 0.00 & 5.56 &
0.00 \\ \hline
  PTU2009W & 4.76 & 23.81 & 4.76 & 9.52 & 9.52 & 4.76 & 19.05 & 0.00 & 23.81 &
0.00 \\ \hline
  PTU2009S & 16.67 & 22.22 & 11.11 & 5.56 & 27.78 & 5.56 & 5.56 & 0.00 & 5.56 &
0.00 \\ \hline
  DBATU\newline2015S & 18.18 & 27.27 & 0.00 & 9.09 & 27.27 & 0.00 & 9.09 & 0.00
& 9.09 & 0.00 \\ \hline
  DBATU\newline2014W & 12.50 & 25.00 & 0.00 & 6.25 & 43.75 & 6.25 & 6.25 & 0.00
& 0.00 & 0.00 \\ \hline
  DBATU\newline2014S & 11.76 & 11.76 & 0.00 & 23.53 & 35.29 & 5.88 & 11.76 &
0.00 & 0.00 & 0.00 \\ \hline

\end{tabular}
\end{table}
\section{Results of the Analysis}
This section describes the results of the analysis carried out 
after the assignment of various question items.
\subsubsection*{Cognitive Level Categorization}
\begin{figure}[]
  \centering
  \includegraphics[scale=0.75]{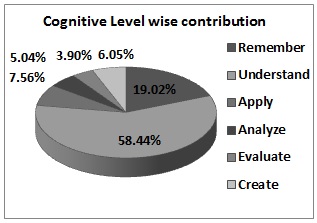}
  \caption{cognitive level wise contribution} \label{cref}
\end{figure}

Table \ref{ca} shows paper-wise analysis of question items as per the
cognitive levels. Entries in the Table \ref{ca} indicate percentage
of the questions that belong to one particular cognitive category. 
For example, in Table  
\ref{ca}, 6.25\% questions are asked  to test the students for the skill of 
$Remeber$ in an examination with the paper ID $MU2014S$. All the paper-wise
cognitive analyses are merged to find the average values for the cognitive categorization
as shown in Figure \ref{cref}. In summary,   students are tested 
for cognitive categories in the order of Understand (58.44\%), Remember(19.02\%),
Apply(7.56\%), Create (6.05\%), Analyze (5.04\%), Evaluate (3.90\%).

\subsubsection*{Knowledge Level Categorization}
\begin{figure}[]
\centering
  \includegraphics[scale=0.75]{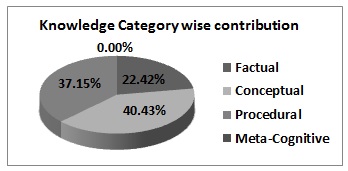}\\
  \caption{Knowledge category  wise contribution} \label{kref}

\end{figure}

Table \ref{kl} shows paper-wise analysis of question items according to
knowledge types. Entries in the Table \ref{kl} indicate percentage
of the questions that belong to one particular type of knowledge.
For example, in Table  
\ref{kl}, 31.25\% questions are asked  to test the students for the $Procedure$ type of knowledge 
 in an examination with the paper ID $MU2014S$. All the paper-wise
knowledge level analyses are merged to find the average values level distribution
as shown in Figure \ref{kref}. In general,  Indian universities test students
for types of knowledge in the order of  Conceptual (40.43\%), Procedural(37.15\%)
and Factual(22.42\%). 

\subsubsection*{Distribution across the Knowledge Areas}
\begin{figure}[]
    \centering
    \includegraphics[scale=0.4]{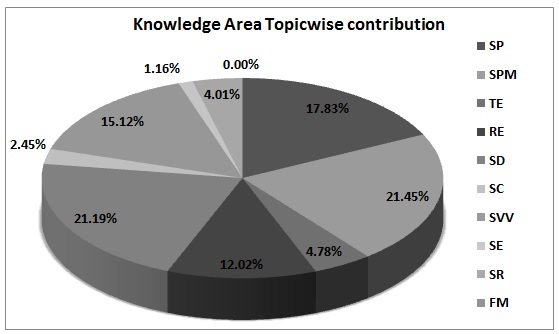}\\
    \caption{Knowledge Unit wise contribution} \label{kuref}
 \end{figure}

Table \ref{ku} shows paper-wise analysis of question items
distributed across the knowledge units. Entries in the Table \ref{ku} indicate percentage
of the questions that belong to one particular knowledge unit. 
For example, in Table  
\ref{ku}, 25\% questions are asked  to test the students for the unit on
$Software\ Design (SD)$ in an examination with the paper ID $MU2014S$. All the paper-wise
analyses are merged to find the average values for distribution of question items across
various knowledge units
as shown in Figure \ref{kuref}. In general, 
Software Design (SD), Software Project Management and Software Processes
are three most favored knowledge units to test
software engineering specific skills. 
Surprisingly no university tests their students for knowledge 
of $Formal Methods(FM)$ in a course on Software Engineering.

\section{Related Work}

We have presented a survey on  skill and knowledge levels  assessed through
software engineering examinations conducted by Indian universities. 
Categories from the Revised Bloom's Taxonomy are used to perform
the analysis of question items. To the best of our knowledge,
this might be  the first attempt of conducting
such a kind of survey in the context of Indian Universities. 
However, the RBT has been extensively applied by earlier researchers
for various purposes. In this section, we present a brief review of
applications of RBT in software engineering education and 
its applications in conducting examinations.

In \cite{boon2014examination}, authors propose a question paper
preparation system based on content-style separation principles. The purpose of
the system was to generate questions  belonging to different categories of 
the Bloom's taxonomy. A generic visual model for an
automated examination  system has been proposed in \cite{naeem2014improving}
using UML as a modeling language. 
The system is generic in the sense that 
it can be configured  according to the requirements of an institution.
Furthermore, the model provides  performance analysis of students.
The authors\cite{whalley2006australasian} present a report on a
multi-institutional investigation into the reading and comprehension skills of
novice programmers. The Bloom's and SOLO
taxonomies  are used to analyze the results of a programming exercises  carried
out by students at a number of universities. 
A rule-based classification scheme to analyze question items using Bloom's taxonomy
is presented in \cite{nzila}. The authors pointed out that 
effectiveness of such classifier systems is one of the concerns while classifying question items
according to Bloom's Taxonomy.

Unlike these earlier  applications of RBT, in this paper,
we combine  RBT and software engineering specific knowledge 
areas and use it as the framework to analyze question items. 
By adding SE knowledge areas in RBT, the analysis framework becomes
more relevant to assess software engineering specific skills.

\section{Conclusion}

The paper presents a qualitative assessment of question items collected from
end semester examinations for  the course on Software Engineering conducted
by various Indian Universities. 
While analyzing the question items,  some of the challenges  relate
with the use of tools and the action-verbs list used during cognitive categorization.
Some  action verbs  appear in more than  than one category.
For example,  the action verb $Choose$ appears in  categories: {\em Remember, Apply, Evaluate},
and {\em Create}.  So, it  becomes difficult to categorize  question items
only on the basis of action verbs. In such situations,  the context of a question
 needs to be taken into consideration for the appropriate categorization of
the question item. 

Combining the RBT framework with domain specific knowledge areas is the main highlight of the analysis
method used in this paper. 
We found  that the Revised Bloom's Taxonomy (RBT)
is a useful framework to assess  generic
skills and knowledge levels tested. But it is inadequate to test domain specific skills in general and
Software Engineering specific skills in particular. 
To overcome this limitation of RBT framework, we extended it  by adding
Software Engineering specific knowledge areas. 
The second highlight of the paper is  the creation  of 
a classified question bank of about eight hundreds questions from the discipline of software engineering.
This question bank in which each question item is classified as per cognitive and knowledge
categories can also be used to test the performance and effectiveness of any automated
tool implemented for categorization of question items

The  results of the analyses  presented in this paper can be used to design an advanced course on Software Engineering by
universities or to design pre-induction training programs by software development organizations.

\bibliographystyle{plain}
\bibliography{btforcs}
\end{document}